\documentclass[10pt]{iopart}

\usepackage{graphicx}
\usepackage{epsfig}
\usepackage{bm}

\def\be{\begin{equation}}
\def\ee{\end{equation}}
\def\bee{\begin{eqnarray}}
\def\eee{\end{eqnarray}}

\begin{document}

\title[Non-linear evolution of the tearing mode using gyro kinetic simulations.]{The non-linear evolution of the tearing mode in electromagnetic turbulence using gyrokinetic simulations.}
 \author{W A Hornsby$^{1,2}$, P Migliano$^1$, R Buchholz$^1$, S. Grosshauser$^1$, A Weikl$^1$, D Zarzoso$^{4,2}$, F J Casson$^3$, E Poli$^2$, A G Peeters$^1$}
\address{$^1$ Theoretical Physics V, Dept. of Physics, Universitaet Bayreuth, Bayreuth, Germany, D-95447}
\address{$^2$ Max-Planck-Institut f\" ur Plasmaphysik, Boltzmannstrasse 2, D-85748}
\address{$^3$ CCFE, Culham Science Centre, Abingdon, Oxon., OX14 3DB, UK}
\address{$^4$ Aix-Marseille Universit\'{e}, CNRS, PIIM, UMR 7345 Marseille, France}

\ead{william.hornsby@ipp.mpg.de}

\date{\today}

\begin{abstract}
The non-linear evolution of a magnetic island is studied using the Vlasov gyro-kinetic code GKW.  The interaction of electromagnetic turbulence with a self-consistently growing magnetic island, generated by a tearing unstable $\Delta' > 0$ current profile, is considered.   The turbulence is able to seed the magnetic island and bypass the linear growth phase by generating structures that are approximately an ion gyro-radius in width.  The non-linear evolution of the island width and its rotation frequency, after this seeding phase, is found to be modified and is dependent on the value of the plasma beta and equilibrium pressure gradients.  At low values of beta the island evolves largely independent of the turbulence, while at higher values the interaction has a dramatic effect on island growth, causing the island to grow exponentially at the growth rate of its linear phase, even though the island is larger than linear theory validity.  The turbulence forces the island to rotate in the ion-diamagnetic direction as opposed to the electron diamagnetic direction in which it rotates when no turbulence is present.  In addition, it is found that the mode rotation slows as the island grows in size.

\end{abstract}

\section{Introduction.}

Magnetic islands in a tokamak can lead to loss of confinement or even major disruptions of plasma.  The tearing mode \cite{FUR63,FUR73}, specifically the neoclassical tearing mode (NTM) \cite{CAR86} is expected to set the beta limit in a reactor \cite{SAU97,WAE09,ITER03}.  

The tearing mode is centred around flux surfaces with rational values of the safety factor.  Current perturbations at these surfaces, coupled with non-ideal effects such as electron inertia or resistivity, can become unstable.   This process, via magnetic reconnection,  forms a growing magnetic island with its corresponding radial magnetic field component.  When collisions can be neglected, the mechanism of reconnection is through the electron inertia, where the singular layer width (the layer where magnetic reconnection can occur), and in turn the growth rate, are related closely to the electron skin depth \cite{HAZ75}, this corresponds to the collision-less tearing mode \cite{DRA77}. When collisions become significant, it is plasma resistivity which determines the singular layer width, producing a classical collisional tearing \cite{FUR63} or, more relevant for present day tokamaks, a semi-collisional mode \cite{DRA77,Fitz10,Ade91}.   In high temperature, weakly collisional, plasmas the mode grows on a slow time-scale. 

The stability and evolution of magnetic islands is the result of the interplay of variety of processes and their interaction with electromagnetic turbulence is further expected to modify their stability and evolution \cite{MIL09}.  Turbulence also generates, and is regulated by zonal and meso-scale flows \cite{ter00}, a process which is further complicated by the presence of magnetic islands \cite{walwael,Hor10,HorVor,POL09,ish07}.   Plasma turbulence, zonal flows and tearing modes occupy disparate time and length scales, however, early in their evolution magnetic islands can be very narrow and thus comparable to turbulent length scales,  as such, their evolution can not be considered to be independent of the turbulence \cite{MIL09,Mur11}, and has been shown to be influenced by turbulence in both gyro-fluid simulations\cite{POL10,ish10} and gyro-kinetic simulations in toroidal geometry \cite{HorNL15}.

Turbulence has been suggested as a source of seed islands for neoclassical tearing modes which display no clear threshold behaviour (i.e they only grow after an initial perturbation is formed) \cite{gude99}, and can also be triggered by an island generated by an unstable classical tearing mode  \cite{reim02}.

The large separation in time-scales, overlap of length scales and the complexity of the multitude of interactions makes analytical theory difficult, therefore here we approach the problem using massively parallel, state-of-the-art kinetic simulation.   
The rest of this paper is organised as follows.  In Sec.~\ref{parameters} we outline the simulation set-up and the parameters considered.  In Sec.~\ref{nonlinear} the non-linear evolution of the tearing mode is described without turbulence present.   In Sec.~\ref{withturb} the non-linear growth of the island self-consistently interacting with electromagnetic turbulence at various values of  $\beta_{e}$ is considered ($\beta_{e}$ is the ratio of the plasma pressure to the magnetic pressure).  The effects of turbulence on the island rotation frequency is considered in Sec.~\ref{rotation}.  Finally, the results are summarised in Sec.~\ref{concs}.

\section{Parameters and set-up.}
\label{parameters}

The global version of the electromagnetic non-linear gyro-kinetic code \small{GKW} \cite{PEE09} was utilised in this study.  A full description of the equations solved, and the implementation of the tearing mode current drive can be found in references \cite{HorLin14,HorNL15}.

For all the simulations presented here the normalised gyro-radius  $\rho_{*} = \rho_{i}/R =  0.005$ is used.  This gives a $n=1$ toroidal mode number with normalised binormal wave-number, $k_{\zeta}\rho_{i} =  0.053$.  In a full tearing mode and turbulence simulation a  total of 36 toroidal modes were used giving a maximum $k_{\zeta}\rho_{i} =  1.86$. The ion and electron temperatures are
assumed to be equal, $T_{i} = T_{e}$.  Resolutions in the parallel, parallel velocity, magnetic moment, radial directions are, $N_{s}=64$, $N_{v}=64$, $N_{\mu}=16$, $N_{x}=256$ respectively.  
This radial resolution places at least three radial grid points per singular layer width depending on the value of $\beta_{e}$.  The linearly growing tearing mode is found to be a semi-collisional mode \cite{DRA77}.  As such, the singular layer width is closely related to the electron skin depth, which is given by,
\begin{math}
\frac{\delta_{e}}{a} = \frac{\rho_{*}}{\sqrt{\frac{m_i}{m_e}}\sqrt{\beta_{e}}}\frac{R}{a}
\end{math}
which must be resolved.
The $\beta_{e}$ is the electron $\beta$ defined as $\beta_{e} = n_{e}T_{e}/(B_{0}^{2}/2\mu_{0})$,  where $n_{e}$ and $T_{e}$ are equilibrium density and
temperature and $B_{0}$ is the magnetic field strength at the outboard mid-plane magnetic axis. 

\begin{figure}
\begin{centering}
\includegraphics[width=9.0cm,clip]{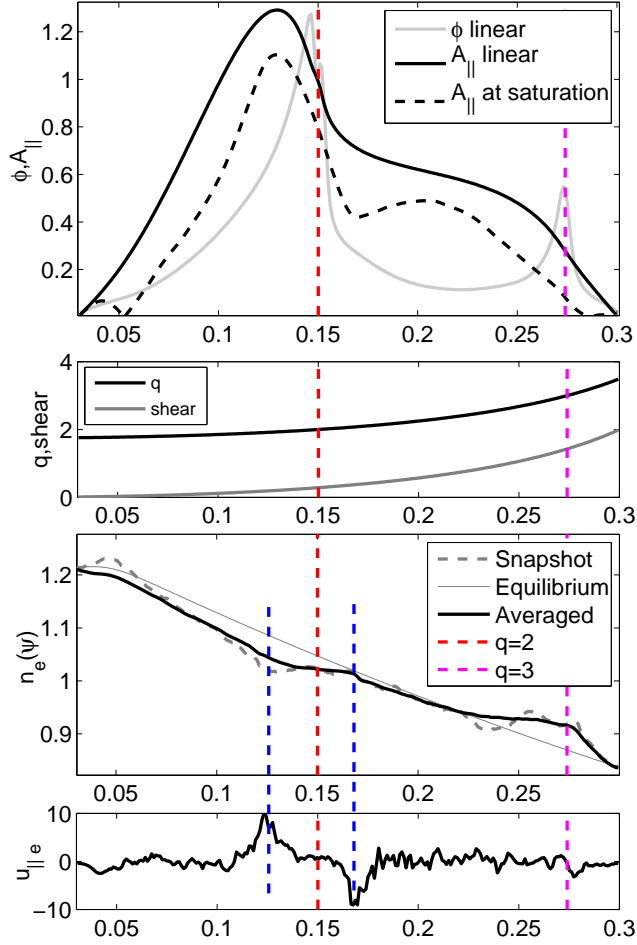}
\caption{The radial profiles of the (top) linear $A_{||}$ and $\phi$ eigenfunctions at $\beta_{e}=0.05\%$ and the $A_{||}$ profile at non-linear saturation (dashed line).  (Upper middle) the safety factor (q) and magnetic shear ($\hat{s}$) profiles. (Bottom middle) the (grey solid) equilibrium electron density profile (black) the time averaged electron density profile through the island O-point and (dashed) a snapshot electron density profile. (Bottom) the perturbed electron current profile at island saturation.  (Blue) vertical dashed lines represent the approximate location of the island seperatrix. }
\label{isleigen}
\end{centering}
\end{figure}

The tearing mode instability is driven by a non-homogeneous current density profile.  In our implementation this 
is introduced by applying an electron flow profile in the equilibrium which is calculated to be self-consistent with the q-profile.    The current profile used in this paper is the same model as used by Wesson et. al. \cite{Wess,Has77}.  Here the current density profile has a parabolic form, $j=j_{0}(1+(r/a)^{2})$, which introduces an electron flow which enters the equations via the source term, 
\begin{equation}
\nabla F_{Me} = -2 \frac{v_{\parallel}}{v_{th}^{2}} \nabla u_{e} F_{Me},
\label{drive}
\end{equation}
where $u_e = u_{e}(r)$ is the electron flow which is related to the current density profile by, $j = n_{e}u_{e}e$.
In the rest of this paper, when a simulation without the drive is mentioned, it is terms of this form of Eq.~\ref{drive} that are neglected from the equations solved.
The safety factor profile is calculated from the current profile, which has the analytic form,
\begin{equation}
q = q_{a}\frac{B_\phi}{R}\frac{r^{2}/a^{2}}{1 - (1 - r^{2}/a^{2})^{2}}
\end{equation}
where $q_{a}$ is the value of the safety factor at radial co-ordinate $\psi = a/R$.  An example of the safety-factor and magnetic shear ($\hat{s}=r/q\partial q/\partial r$) used can be seen in the middle panel of Fig.~\ref{isleigen}.   In the simulations presented here, $q$, at the at the plasma edge is set to $q_{a} = 3.5$ and the safety on the axis is $q(0)=1.75$.  Therefore,  $q=2$, $q=2.5$, $q=3$ and many more higher mode number rational surfaces ($q=m/n$, where m is the poloidal
mode number) exist within our computational domain.   The radial domain covers $0.1a \leq r \leq a$ where a is the minor radius.

The safety factor profile given above is found to provide a linearly unstable semi-collisional tearing mode \cite{HorLin14}, with a positive stability parameter \cite{FUR63} defined as, 
\begin{equation}
\left.\Delta' = \frac{1}{A_{\parallel}}\frac{\partial A_{\parallel}}{\partial r}\right|^{r_{s}^{+}}_{r_{s}^{-}}
\end{equation}
across the singular layer, whose position is denoted by $r_{s}$ and upper and lower boundaries by $r_{s}^{+}$ and $r_{s}^{-}$ .  This parameter represents the discontinuity in $\partial A_{||}/\partial r$ and measures whether mode growth is energetically favourable ($\Delta' > 0$ for the mode to grow).
The tearing mode has the familiar \cite{NISH98} linear structure of which is shown in Fig.~\ref{isleigen}, showing the position of the resistive layers at the $q=2$ and $q=3$ rational surfaces. 

While a full Fokker-Planck collision operator is implemented in GKW, here we use only pitch-angle scattering of the electrons off the bulk ions for simplicity.  The normalised collision frequency (normalised to the trapping/de-trapping rate), $\nu_{*} = 4\nu_{ei}/3\sqrt(\pi\epsilon^{3}) = 0.12$, where  the collision frequency is defined as
$\nu_{ei} = \frac{n_{i}e^{4}\log{\Lambda_{ei}}}{4\pi\epsilon_{0}^{2}m_{e}^{2}v_{e}^{3}}$.   The density and temperature profiles have the radial form,
\begin{eqnarray}
\frac{\partial n_{s}}{\partial r} = \frac{1}{2}\frac{R}{L_{ns}}(\tanh{(r-r_{0}+\Delta r)} -\tanh{(r-r_{0}-\Delta r)})\nonumber\\
\frac{\partial T_{s}}{\partial r} = \frac{1}{2}\frac{R}{L_{Ts}}(\tanh{(r-r_{0}+\Delta r)} -\tanh{(r-r_{0}-\Delta r)})\nonumber
\end{eqnarray}
where $R/L_{n} = -(R/n)\partial n/\partial r$ and $R/L_{T} = -(R/T)\partial T/\partial r$ are the logarithmic gradients, $\Delta r$ is the length scale on which the gradient increases from zero to this at the boundaries.  This form maintains a constant gradient across most of the radial domain, while dropping the gradients to zero close to the boundaries to prevent numerical issues.   In all cases the simulation was initialised with grid-scale noise at a small amplitude.

\section{Non-linear evolution and saturation}
\label{nonlinear}

\begin{figure*}
\begin{centering}
 \includegraphics[width=14.5cm,clip]{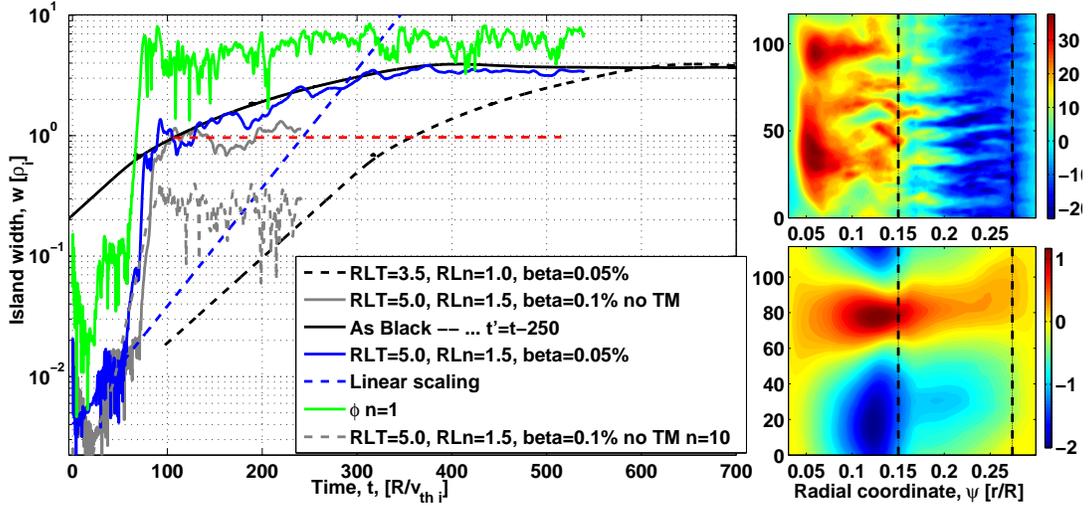}
\caption{(left) The time evolution of parallel vector potential amplitude ($A_{||}$) at the $q=2$ resonant surface for (black dashed) $\beta_{e}=0.1\%$ with $R/L_n = 1.0$ and  $R/L_T = 3.5$,  (blue)  $\beta_{e}=0.05\%$ with $R/L_n = 1.5$ and  $R/L_T = 5.0$ the dashed line represents the linear scaling of the unstable TM with these parameters.  The solid black line is the same simulation as the dashed black line, but transposed so that the amplitudes match at the end of the island seeding phase.
(right) A snapshot of (top) $\phi$ and (bottom) $A_{||}$ at island saturation.  Vertical dashed lines correspond to the radial positions of the singular layers.}
\label{lowbetatrace}
\end{centering}
\end{figure*}

Figure \ref{lowbetatrace} shows (black dashed) the time evolution of the magnetic island width for a non-linear simulation where the background pressure gradient is insufficiently large to drive the ITG mode unstable.   The magnetic island width is related to the magnitude of the vector potential (in the constant-$\psi$ approximation) by the expression
\begin{equation}
w = \rho_{i}\sqrt{\frac{4qA_{\parallel 0}}{\hat{s}}}
\end{equation}  
where $A_{\parallel 0}$ is the mode amplitude at the singular surface.

The parameters utilised were  $R/L_{n} = 1.0$, $R/L_{T} = 3.5$ and $\beta_{e}=0.1\%$.  In this case the island evolves in the manner as predicted by Rutherford \cite{RUTH73}.  At small time, the island grows exponentially at its linear growth rate until it reaches a width (at $\sim 300 R/v_{th i}$) that is approximately the singular layer width ($\sim 0.7\rho_{i}$ Parameters are shown in Table \ref{growths}, Case 2.).  After this the growth of the island slows down and enters the non-linear regime \cite{RUTH73} where the growth of the island becomes algebraic rather then exponential ($w \propto t$).  Eventually the island reaches a saturated width which is calculated here to be $3.7\rho_{i}$,  consistent with quasilinear calculations for this form of equilibrium current profile \cite{White77}.  Also shown in Fig.~\ref{isleigen} is the electron density profile, where the flattening of the profile through the O-point is evident at both the $q=2$ ($m/n=2/1$ mode) surface and the smaller island at the $q=3$ ($m/n=3/1$) surface.

\begin{figure*}
\begin{centering}
 \includegraphics[width=13.5cm,clip]{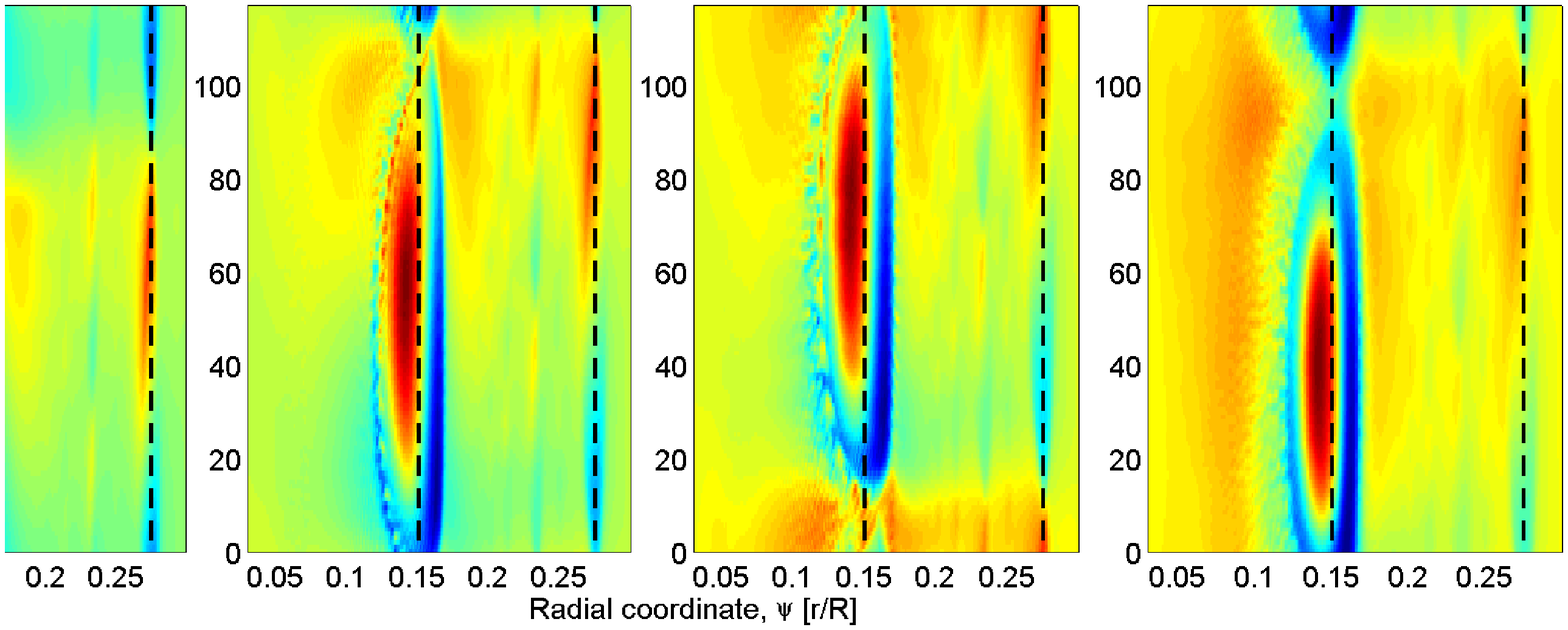}
 \includegraphics[width=5.5cm,clip]{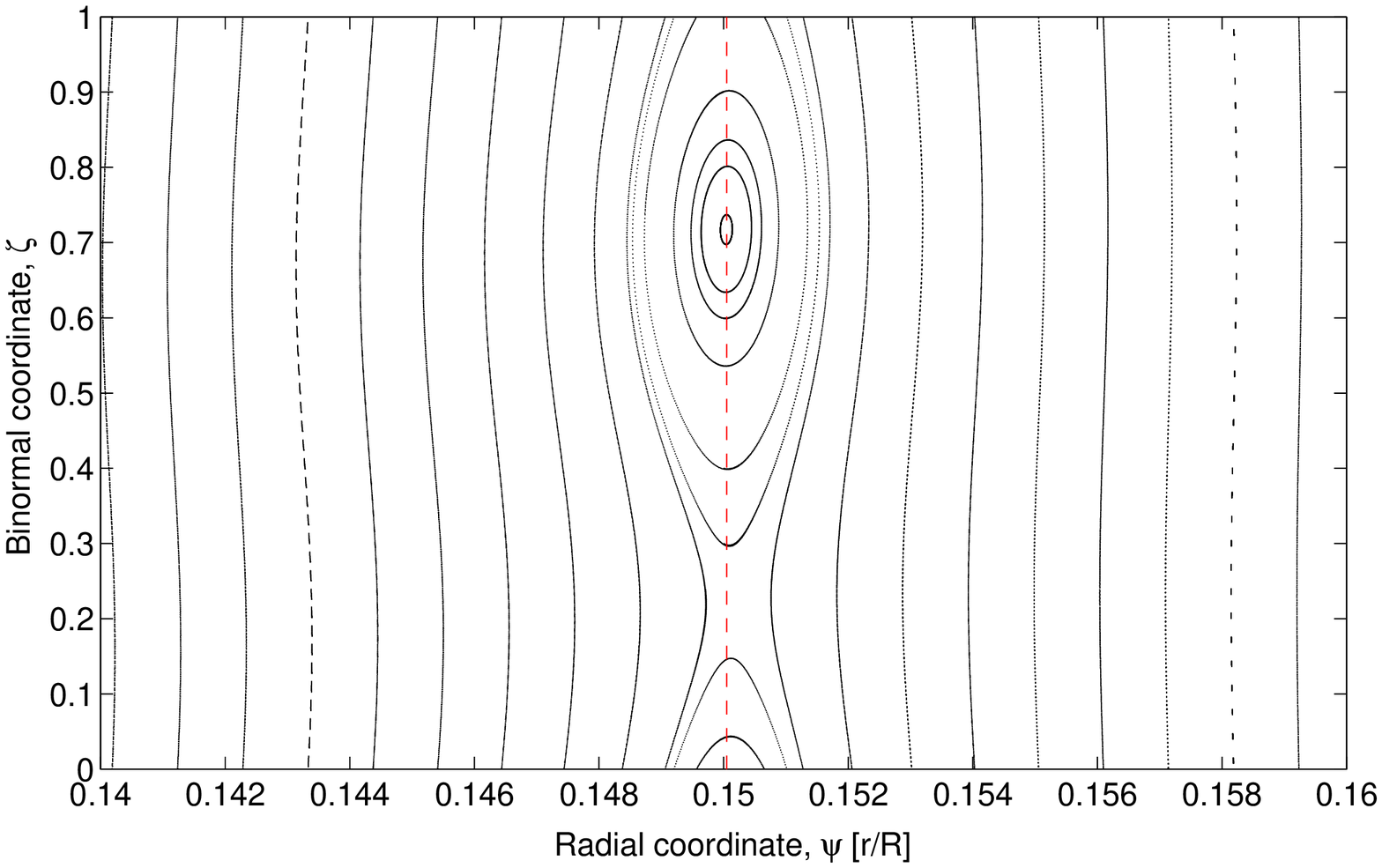}
 \includegraphics[width=5.4cm,clip]{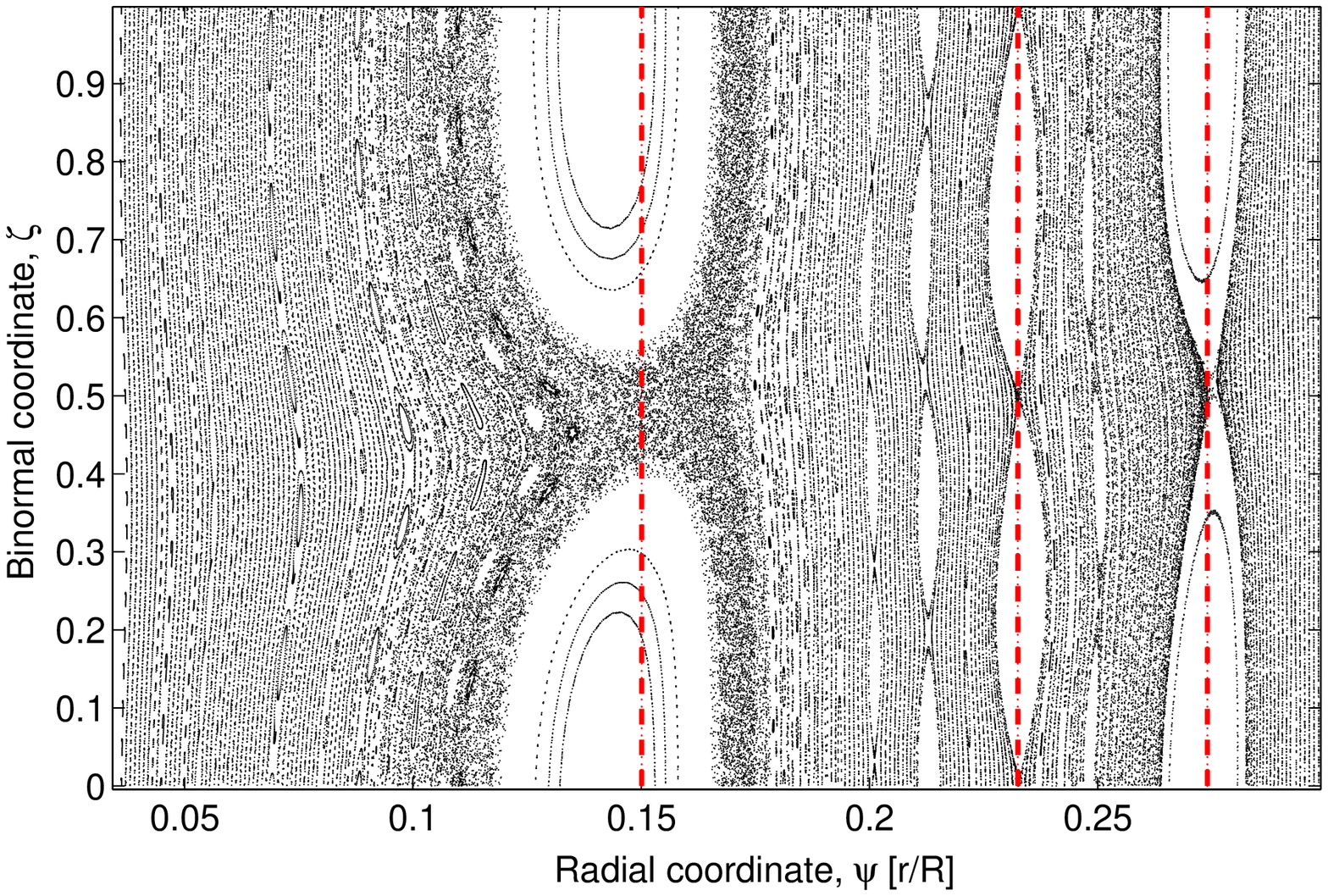}
\caption{Snapshots of the non-linear evolution of the electro-static potential of a growing magnetic island when the equilibrium gradients are insufficient to drive micro-instabilities.  However, the growing island can induce a small scale drift-wave around its separatrix, causing a stochastisation around the seperatrix.  This stochastisation can be seen in the Poincar\'{e} plot (bottom right) which corresponds to the last time slice at island saturation, (bottom left) is the same but during the linear phase of the magnetic island evolution.  The $q=2$, $q=2.5$ and $q=3$ resonant surfaces are also plotted and higher order islands are also visible.}
\label{smallitgisland}
\end{centering}
\end{figure*}

Stochastisation of magnetic islands is a well known phenomenon, and it has also been shown in non-linear gyro-kinetic simulations of magnetic islands with turbulence that higher mode numbers may excite and cause stochastisation of the island seperatrix and an almost complete break down of the island X-points \cite{HorNL15}.  Here however it can be seen that even when the equilibrium gradients are too small for turbulence to form, the non-linear modification of the density and temperature profiles can cause a steepening of gradients near the island seperatrix (also shown in Fig.~\ref{isleigen}) and this, in turn, can cause a high-k ITG to become unstable in the island vicinity.  Figure \ref{smallitgisland}, shows four time slices of the island electro-static potential during the non-linear phase of the island evolution.  It is evident that a high mode number, drift-wave becomes unstable at the island seperatrix.  A similar process has been observed in 2D fluid simulations of an tearing mode \cite{Hu14}.   These small scale ITG modes can then interact and form turbulence, which in turn can cause the island seperatrix to become stochastised.  The bottom right panel shows a Poincar\'{e} plot of the magnetic field at the saturated state, clearly showing a stochastic layer around the island which, initially, had a coherent form (bottom left).  Also visible are islands generated at higher order mode numbers (e.g. m/n=8/3, 9/4) that are linearly stable, but here are non-linearly excited and causing stochastisation from island overlap \cite{bick,lich}.

\section{Turbulence interaction}
\label{withturb}

As the equilibrium density and temperature gradients are increased, the ITG becomes unstable and electromagnetic turbulence forms.  A snapshot of a $\psi-\zeta$ plane during a turbulence simulation is shown in the right hand panels of Fig.~\ref{lowbetatrace}. They show the electro-static potential (top) and electromagnetic potential (bottom) at point where the island saturates $t\sim 700 R/v_{thi}$.  The turbulence, in turn, interacts with a growing magnetic island.

\begin{table*}
\centering
\begin{tabular}{|p{1.0cm}||p{1.0cm}||p{1.0cm}||p{1.0cm}||p{2cm}||p{1.7cm}||p{1.7cm}|}
\hline
Case & $\beta_{e}$ ($\%$) & $R/L_n$ & $R/L_T$ & Growth rate, $\gamma$ ($v_{th}/R$) & Frequency, $\omega$ ($v_{th}/R$) & Electron skin depth ($\rho_i$)\\
\hline
1 & 0.05 & 1.5 & 5.0 & 0.045 & -0.16 & 1.04\\
2 & 0.1 & 1.0 & 3.5 & 0.032 & -0.12 & 0.73 \\
3 & 0.1 & 1.5 & 5.0 & 0.024 & -0.15 & 0.73 \\
4 & 0.2 & 1.5 & 5.0 & 0.0090 & -0.15 & 0.52\\
5 & 0.1 & 2.2 & 6.9 & 0.0032 & -0.28 & 0.73\\
\hline
\end{tabular}
\caption{Table of the growth rates, mode frequencies and the corresponding electron skin-depths for the corresponding parameters in simulations utilised in this paper.  All linear calculations are for the $n=1$ toroidal mode, and the most unstable mode is always found to be a tearing mode.}
\label{growths}
\end{table*}

Firstly we concentrate on a simulation with an electron $\beta_{e}=0.05\%$, (Case 1 in Table \ref{growths}) here the turbulence is fundamentally electro-static, and the tearing mode has a large singular layer width (Since $\delta_{e}\sim 1/\sqrt{\beta_{e}} \sim \rho_{i}$). The equilibrium temperature and density gradients have the logarithmic gradients, $R/L_{T}=5.0$ and $R/L_{n}=1.5$ respectively.   Gradients are at the reference radius, $r = r_{0} = 0.5a$ and are the same for both ions and electrons.  These gradients are large enough to drive micro-instabilities which lead to turbulence but not large enough to completely stabilise the linear tearing mode \cite{Dra83,HorLin14}, an observation that was confirmed by linear simulations of the $n=1$ toroidal mode using the same parameters.  The calculated growth rates and frequencies are shown in Table \ref{growths}.   

The time trace of the $n=1$ toroidal component of the electro-magnetic potential ($A_{||}$) for these simulations are shown in figure ~\ref{lowbetatrace}.  Also included is the trace of the $n=1$ electro-static potential, $\phi$ (Green line)  showing that it is largely determined by the turbulence, reaching rapidly a saturated state.  In contrast, the island is still growing as evidenced by the increasing $A_{||}$, the electrostatic potential, $\phi$, has saturated.  It can be seen comparing the trace with turbulence (blue line) to that without (black dashed line), that the turbulence provides an initial island structure \cite{ish10,Poye15,itoh04} that is approximately the size of the ion-gyro-radius.  A gyro-kinetic study of this seeding mechanism can be found in \cite{HorNL15}.  Without a background current profile (grey line) the amplitude of the mode saturates and maintains this amplitude, but with a tearing unstable current profile the seed island structure starts to grow (from approximately $t=100 R/v_{th i}$) and eventually saturates at a half width that is calculated to be $w = 3.7\rho_{i}$ after $700 R/v_{th i}$.

Taking the curve of the island evolution without turbulence present, and shifting the x-axis so that the mode amplitudes are comparable at the end of the seeding phase (approx. $100 R/v_{th i}$).  It can be seen that the non-linear island evolution after this point agree almost exactly from this point until island saturation.  The saturation amplitudes also have a very close agreement.  While the equilibrium pressure gradients in these two simulations are slightly different, giving different linear growth rates, this is not expected to change significantly the non-linear behaviour of the island, therefore  a comparison can  be made.  It is seen that the turbulence seeding mechanism bypasses the linear regime entirely, providing a structure much bigger than the singular layer width in almost all the cases considered, and approximately the layer width in this low beta case.  Above this width the island evolves, with this value of $\beta_{e}$, almost independent of the turbulence around it and is governed by the positive stability parameter $\Delta'$.
                                   
\begin{figure}[h!]
\begin{centering}
 \includegraphics[width=8.5cm,clip]{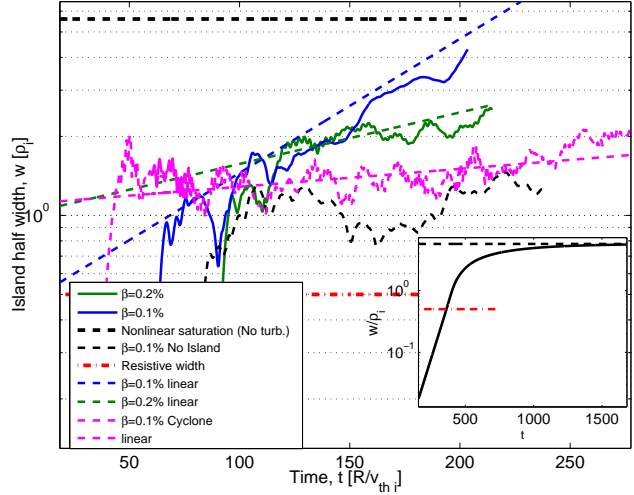}
\caption{The time evolution of the $n=1$ parallel vector potential $A_{||}$ for four different simulations where the equilibrium temperature gradient, density gradients and electron $\beta_{e}$ are varied.  Values are (blue) $\beta_{e}=0.1\%$, $R/L_{T}=5.0$, $R/L_{n}=1.5$ and (green) $\beta_{e}=0.2\%$, $R/L_{T}=5.0$, $R/L_{n}=1.5$ and for (magenta) $\beta_{e}=0.1\%$, $R/L_{T}=6.9$, $R/L_{n}=2.2$.  (black) uses the same parameters as the blue traces but without a current profile.  The dashed lines are the corresponding linear island growth scalings.  Horizonal (black) dashed and (red) dashed lines are the island saturated width and singular layer width respectively from a simulation with only two toroidal modes (shown in inlay). }
\label{islandturbtrace}
\end{centering}
\end{figure}

However, as $\beta_{e}$ is increased, a marked change in the island evolution is observed.  Figure \ref{islandturbtrace} shows the time evolution of the island width in simulations where $\beta_{e}$ was increased to $0.1-0.2\%$ (Cases 3, 4 and 5 in Table \ref{growths}).  At early time a similar evolution to the low $\beta_{e}$ case is seen.  The electromagnetic component of the turbulence provides a seed magnetic island structure which further grows, once turbulence has saturated, if there is a tearing unstable current profile present.  However, after this point a different evolution is seen.  Without turbulence the island evolves non-linearly with an algebraic growth, $w \propto t$.  With turbulence, in all three simulations presented here varying $\beta_{e}$ and equilibrium pressure gradients, the island grows exponentially at a rate that is the same as its corresponding linear growth rate.  

The linear growth scaling of these simulations are shown as corresponding colour dashed lines in the figure.  The calculated growth rates and frequencies are outlined in Table \ref{growths}.  Two stabilisation effects are evident.  Firstly, as the density and temperature gradients are increased (for constant $\beta_{e}$) the linear growth rate drops due to the generation of diamagnetic currents outside the singular layer, which cause a stabilisation of the tearing mode \cite{Dra83}.   Pressure gradients, combined with magnetic curvature are also known to have a stabilising effect on the linear \cite{GGJ75} and non-linear mode growth \cite{kotsch85}, however at these low $\beta_{e}$, the effects are likely to be small.   Secondly, as the $\beta_{e}$ is increased, the growth rate reduces due to the narrowing of the singular layer width.  
Drake and Lee showed that for a collisionless tearing mode (semi-collisional is a slight modification of this) that the growth rate $\gamma \sim \delta_{e}^{2}$ where $\delta_{e}$ is the skin depth.   A higher $\beta_{e}$ reduces the skin depth and thus the growth rate, at lower $\beta_{e}$ the skin depth is wider and thus a larger growth rate is seen.

In all cases the electromagnetic turbulence supplies a seed island structure of approximately an ion gyro-radius in width, greater than or close to the size of the singular layer width (Values are also outlined in Table \ref{growths}) and thus linear growth scaling should not be valid here.  The island, when turbulence is not present or in the lower $\beta_{e}$ case seen in the previous section, at this size grows algebraically.  However, the electromagnetic component of the turbulence here is larger and thus a stronger interaction between the island, turbulence and its corresponding zonal flow is evident \cite{ish07,ish10}.  This interaction has the effect of disrupting or modifying the threshold of the non-linear mechanism which, without turbulence, would slow the islands evolution.  Thus, due to the interaction of the island with turbulence, the growth rate is significantly faster than expected.

\section{Island rotation.}
\label{rotation}

The amplitude and the sign of the island's rotation frequency, $\omega$, is of direct consequence to models of its non-linear stability \cite{wael01,Pol05,Smol93}.    The polarisation current \cite{WIL96}, one of the mechanisms invoked to explain NTM threshold behaviour \cite{Butt04}, is closely related to the mode frequency.  Whether or not the polarisation current is stabilising is dependent on the sign of the frequency \cite{mikh03,Ima12}.  As such, a knowledge of the islands rotation frequency, particularly when the width is small, is critical in understanding island stability.   Moreover, in the case of finite overlap of the ion orbits with the island, the rotation of the island is known to have an effect on the density profile through the magnetic island, which can enhance or counteract the flattening of the profile depending on its sign \cite{Smol93,Sicc11}.  This effect has been observed also in the presence of turbulence \cite{DZAR15}.  The bootstrap current profile, the drive of the Neoclassical tearing mode \cite{CAR86,Berg09}, is directly related to the density profile.  
 
\begin{figure}
\begin{centering}
 \includegraphics[width=8.5cm,clip]{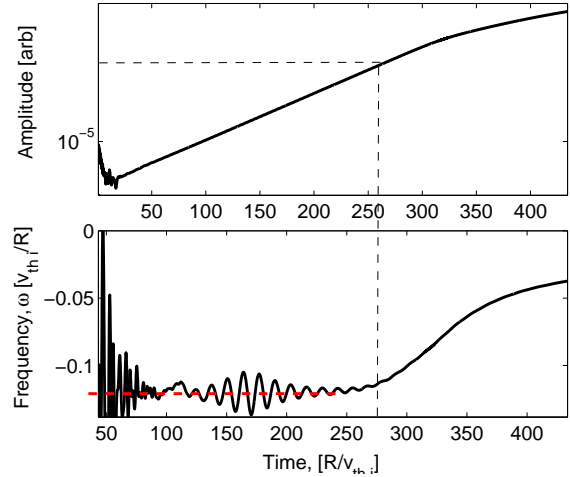}
\caption{(top) Mode amplitude and (bottom) mode frequency for a growing magnetic island in the absence of turbulence.  The black dashed lines denote the approximate amplitude and time where the island starts to evolve non-linearly.  The red horizontal dashed line denotes the linearly calculated mode frequency.}
\label{islandfreqtrace}
\end{centering}
\end{figure}

When turbulence is not present the magnetic island, in the low-collisionality linear regime, rotates at the electron diamagnetic direction \cite{Dra83} and then the rotation frequency reduces in amplitude as the island grows and enters the non-linear regime.  Since toroidal angular momentum is conserved in the absence of external sources, as the island grows the plasma mass within the island increases and thus its frequency must decrease \cite{smol95}.   This point in the simulation is denoted by the grey dashed lines.  This is shown in Fig~\ref{islandfreqtrace} for a simulation of an island without turbulence (Case 3 in Table \ref{growths} ) which shows the mode amplitude (and therefore island width) and its frequency as a function of time.

\begin{figure}
\begin{centering}
\includegraphics[width=8.0cm,clip]{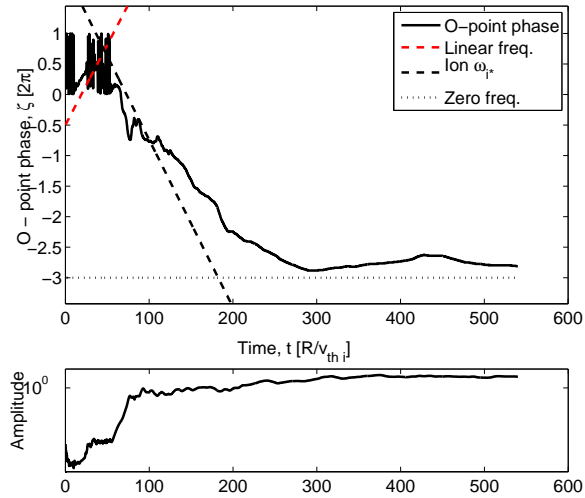}
\caption{(top) The time evolution of the phase of the magnetic island O-point for simulations where electromagnetic turbulence is present in the low $\beta_{e}$ case.  The (red dashed) line represents the scaling with the linear frequency.  (Black dashed) The scaling of the phase with the ion-diamagnetic frequency ($\omega_{i*}$). (bottom) for reference the amplitude of the $n=1$ electromagnetic mode.}
\label{islandotrace}
\end{centering}
\end{figure}

However, the above description is significantly altered when turbulence is present.  Figure \ref{islandotrace} shows the time trace of phase, $\zeta$, of the O-point as a function of time for case 1.  Where the phase is related to the mode frequency by $\omega = -d\zeta/dt$.  This is achieved by knowing that the mode with the longest binormal wave-vector is the most tearing unstable and thus produces the dominant island.  The position of the O-point is found by Fourier filtering away all higher k modes and then finding the minima of the mode in the binormal direction.  This location, and thus the island phase, can then be traced through a turbulent simulation.  For reference, an increasing phase with time is in the electron diamagnetic direction, while the opposite is in the ion direction.

Initially, the $n=1$ mode has a frequency in the electron direction, as expected for a linearly growing tearing mode (rotation at the linear frequency is shown by the red dashed line), or micro-tearing mode that can be found at this mode number when the current drive term is neglected.  Once the ITG grows and turbulence is established it can be seen that the island mode begins to rotate in the ion diamagnetic direction.  The the black dashed line represents the rotation at the ion-diamagnetic frequency ($\omega_{i*}=\frac{k_{\zeta}\rho_{i}}{2}\frac{R}{L_{n_i}}$) of the $n=1$ toroidal mode, and it is evident that the island rotates at approximately this frequency.   However, it is also visible that when there is an island growing, the mode frequency is also a function of the island's width.  The island slowing down as it grows and reaching almost a zero frequency state when it saturates.  It may also, as is visible here for the $\beta_{e}=0.05\%$ case, temporarily change sign but maintaining a small frequency amplitude.

\begin{figure}  
\begin{centering}
 \includegraphics[width=8.5cm,clip]{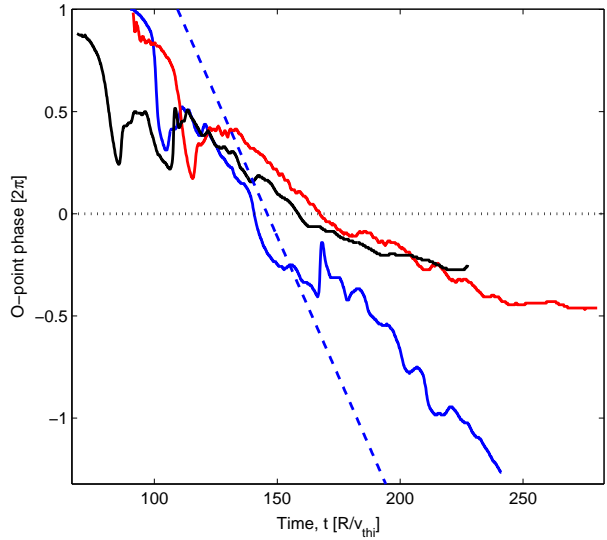}
\caption{The time evolution of the position of the magnetic island O-point for simulations where electromagnetic turbulence is present.  The (blue dashed) line represents the rotation at the ion diamagnetic velocity for the $n=1$ mode.  The blue trace is for a simulation where the island is not growing as no equilibrium current profile is imposed, while the red and black curves represent simulations where magnetic islands are growing from a seed, Cases 4 and 3 respectively.}
\label{islandotrace2}
\end{centering}
\end{figure}

Figure \ref{islandotrace2} also shows the time traces of the island phase for cases 3, 4 and for the same parameters as case 3 but without the current gradient drive term to form a growing island.  Here it can be clearly seen that the rotation frequency of the mode when there is no growing island is approximately constant, while when a growing island is present in the non-linear phase, the frequency is a function of the island width.  The frequency decreases in amplitude as the island grows.  This island deceleration does not seem to change significantly with increasing $\beta_{e}$ or between the different equilibrium gradients between these two simulations.

\section{Conclusions}
\label{concs}

Using state-of-the-art gyro-kinetic simulations  
this paper has presented the non-linear 
evolution of magnetic islands driven unstable by the classical tearing 
mode, and their self-consistent interaction with electromagnetic 
turbulence.

A number of simulations were performed varying the plasma $\beta_{e}$ and the equilibrium temperature and density profiles, but in all cases it was found that the turbulence provides, through non-linear interactions, a seed island structure.  However, how the island evolves after this phase is dependent on the plasma electron $\beta_{e}$.  At lower beta, the island evolves non-linearly largely free of interaction with the turbulence, confirmed by comparison with simulations of an island an the absence of turbulence.  However, at higher values of beta, the interaction becomes stronger and has the effect of causing the island to grow exponentially, even though it is no longer in a regime where this is expected because the island widths are significantly larger than the singular layer width.

A magnetic island will rotate in the electron diamagnetic direction 
when no turbulence is present at low collisionality.  This rotation slows down as the island 
grows, through conservation of momentum, since a growing island captures 
a larger plasma mass as it grows.  However it is observed that, when present, the turbulence forces the magnetic island to rotate in the ion-diamagnetic direction even when the island is small.   When a growing island is present, as it grows larger, the rotation slows down, eventually stopping and periodically changing sign as it approaches a saturated width.

Non-linear simulations with turbulence may excite modes that are linearly tearing stable and form higher mode number islands.  The presence of turbulence and higher order modes can cause the separatrix and X-point to become stochastised.  However, it was also observed that even when the ITG mode is stable, the island's influence on the background equilibria (flattening the profiles within, and steeping just outside) can cause a high mode number ITG to become unstable and drives turbulence to form around the separatrix, which causes a stochastisation of the field lines in this area.  This process could have significant implications for boundary layer theories of tearing stability, as even islands evolving in the presence of small gradients can cause the generation of turbulence and stochastisation.

\ack

A part of this work was carried out using the HELIOS supercomputer system at Computational Simulation Centre of International Fusion Energy Research Centre (IFERC-CSC), Aomori, Japan, under the Broader Approach collaboration
between Euratom and Japan, implemented by Fusion for Energy and JAEA.

This work was partially carried out thanks to the support of the A*MIDEX project (no.
ANR-11-IDEX-0001-02) funded by the ÒInvestissements dÕAvenirÓ French
Government program, managed by the French National Research Agency (ANR) and with funding from the RCUK Energy Programme [grant number EP/I501045].

One of the authors (WAH) would like to thank Clemente Angioni for many useful discussions.

\end{document}